\PassOptionsToPackage{unicode}{hyperref}
\PassOptionsToPackage{hyphens}{url}
\PassOptionsToPackage{dvipsnames,svgnames,x11names}{xcolor}
\documentclass[
  runningheads]{llncs}

\usepackage{amsmath,amssymb}
\usepackage{iftex}
\ifPDFTeX
  \usepackage[T1]{fontenc}
  \usepackage[utf8]{inputenc}
  \usepackage{textcomp} 
\else 
  \usepackage{unicode-math}
  \defaultfontfeatures{Scale=MatchLowercase}
  \defaultfontfeatures[\rmfamily]{Ligatures=TeX,Scale=1}
\fi
\usepackage{lmodern}
\ifPDFTeX\else  
\fi
\IfFileExists{upquote.sty}{\usepackage{upquote}}{}
\IfFileExists{microtype.sty}{
  \usepackage[]{microtype}
  \UseMicrotypeSet[protrusion]{basicmath} 
}{}
\usepackage{xcolor}
\ifx\paragraph\undefined\else
  \let\oldparagraph\paragraph
  \renewcommand{\paragraph}[1]{\oldparagraph{#1}\mbox{}}
\fi
\ifx\subparagraph\undefined\else
  \let\oldsubparagraph\subparagraph
  \renewcommand{\subparagraph}[1]{\oldsubparagraph{#1}\mbox{}}
\fi

\usepackage{longtable,booktabs,array}
\usepackage{calc} 
\usepackage{etoolbox}
\makeatletter
\patchcmd\longtable{\par}{\if@noskipsec\mbox{}\fi\par}{}{}
\makeatother
\IfFileExists{footnotehyper.sty}{\usepackage{footnotehyper}}{\usepackage{footnote}}
\makesavenoteenv{longtable}
\usepackage{graphicx}
\makeatletter
\def\maxwidth{\ifdim\Gin@nat@width>\linewidth\linewidth\else\Gin@nat@width\fi}
\def\maxheight{\ifdim\Gin@nat@height>\textheight\textheight\else\Gin@nat@height\fi}
\makeatother
\setkeys{Gin}{width=\maxwidth,height=\maxheight,keepaspectratio}
\makeatletter
\def\fps@figure{htbp}
\makeatother

\DeclareUnicodeCharacter{03BC}{$\mu$}
\usepackage{mathtools}
\makeatletter
\@ifpackageloaded{caption}{}{\usepackage{caption}}
\AtBeginDocument{%
\ifdefined\contentsname
  \renewcommand*\contentsname{Table of contents}
\else
  \newcommand\contentsname{Table of contents}
\fi
\ifdefined\listfigurename
  \renewcommand*\listfigurename{List of Figures}
\else
  \newcommand\listfigurename{List of Figures}
\fi
\ifdefined\listtablename
  \renewcommand*\listtablename{List of Tables}
\else
  \newcommand\listtablename{List of Tables}
\fi
\ifdefined\figurename
  \renewcommand*\figurename{Figure}
\else
  \newcommand\figurename{Figure}
\fi
\ifdefined\tablename
  \renewcommand*\tablename{Table}
\else
  \newcommand\tablename{Table}
\fi
}
\@ifpackageloaded{float}{}{\usepackage{float}}
\floatstyle{ruled}
\@ifundefined{c@chapter}{\newfloat{codelisting}{h}{lop}}{\newfloat{codelisting}{h}{lop}[chapter]}
\floatname{codelisting}{Listing}

\makeatother
\makeatletter
\@ifpackageloaded{caption}{}{\usepackage{caption}}
\@ifpackageloaded{subcaption}{}{\usepackage{subcaption}}
\makeatother
\ifLuaTeX
  \usepackage{selnolig}  
\fi
\usepackage[style=lncs,maxcitenames=2]{biblatex}
\usepackage{bookmark}

\IfFileExists{xurl.sty}{\usepackage{xurl}}{} 
\hypersetup{
  pdftitle={Optimizing Communication for Latency Sensitive HPC Applications on up to 48 FPGAs Using ACCL},
  pdfauthor={Marius Meyer; Tobias Kenter; Lucian Petrica; Kenneth O'Brien; Michaela Blott; Christian Plessl},
  pdfkeywords={FPGA, HLS, HPC, inter-FPGA Communication},
  colorlinks=true,
  linkcolor={blue},
  filecolor={Maroon},
  citecolor={Blue},
  urlcolor={blue},
  pdfcreator={LaTeX via pandoc}}

\title{Optimizing Communication for Latency Sensitive HPC Applications
on up to 48 FPGAs Using ACCL}
\titlerunning{ACCL HPC Applications}
\authorrunning{Meyer et al.}
\author{
Marius Meyer\inst{1,2}\and 
Tobias Kenter\inst{1,2}\and 
Lucian Petrica\inst{3}\and 
Kenneth O'Brien\inst{3}\and 
Michaela Blott\inst{3}\and 
Christian Plessl\inst{1,2}}
\institute{
Paderborn University%
, Department of Computer Science%
, Warburger Str. 100%
, Paderborn%
, Germany%
\\
\email{\{marius.meyer, kenter, christian.plessl\}@uni-paderborn.de}
\and 
Paderborn Center for Parallel Computing%
, Mersinweg 5%
, Paderborn%
, Germany%
\and 
AMD Research%
, 2020 Bianconi Ave%
, Dublin%
, Ireland%
\\
\email{\{lucian.petrica, ken.obrien, michaela.blott\}@amd.com}
}
\begin{document}
\maketitle

\begin{abstract}
Most FPGA boards in the HPC domain are well-suited for parallel scaling
because of the direct integration of versatile and high-throughput
network ports. However, the utilization of their network capabilities is
often challenging and error-prone because the whole network stack and
communication patterns have to be implemented and managed on the FPGAs.
Also, this approach conceptually involves a trade-off between the
performance potential of improved communication and the impact of
resource consumption for communication infrastructure, since the
utilized resources on the FPGAs could otherwise be used for
computations. In this work, we investigate this trade-off, firstly, by
using synthetic benchmarks to evaluate the different configuration
options of the communication framework ACCL and their impact on
communication latency and throughput. Finally, we use our findings to
implement a shallow water simulation whose scalability heavily depends
on low-latency communication. With a suitable configuration of ACCL,
good scaling behavior can be shown to all 48 FPGAs installed in the
system. Overall, the results show that the availability of inter-FPGA
communication frameworks as well as the configurability of framework and
network stack are crucial to achieve the best application performance
with low latency communication.
\keywords{FPGA \and HLS \and HPC \and inter-FPGA Communication}
\end{abstract}

\section{Introduction}\label{introduction}

ACCL \autocite{accl} is a collective communication library for
field programmable gate arrays (FPGAs) offering an MPI-like API for data
exchange between multiple FPGAs using a packet-switched network
established directly via the QSFP ports of the FPGA boards. These ports
allow data transfer with speeds up to 100 GBit/s while open-source
network stack implementations of UDP \autocite{vnx} and TCP
\autocite{easynet} exist to transfer data via these ports. Without ACCL,
integrating these network stacks into a user design is non-trivial
because it requires the user to have detailed knowledge about the
network protocol itself and its implementation. Also, it requires
implementing common communication patterns from scratch for every
project and handling low-level details like IP addresses or network
sockets.

ACCL aims to provide a higher level of abstraction by implementing a
message-passing interface well-suited for
high performance computing (HPC) applications, which is usable within
high level synthesis (HLS) FPGA applications. It supports
well-established MPI-like communication patterns including
point-to-point communication, collectives, and communicators, while
abstracting away the underlying network stack and networking details. As
a downside, ACCL itself and the network stacks consume additional
hardware resources on the FPGA which limits the resources available for
the actual application. This introduces a trade-off between single-FPGA
application performance and communication latency and throughput, which
again may benefit the overall application performance. As a different
approach -- and on the other side of the discussed trade-off -- HPCC
FPGA \autocite{meyer_multi-fpga_2023} and MVAPICH2-FPGA
\autocite{contini_mpifpga_2023} use the host MPI implementation for
inter-FPGA communication, avoiding additional FPGA resource utilization,
but incurring increased communication latency and limited throughput.

In this work, we evaluate the performance of ACCL and its network stacks
for use in multi-FPGA applications and compare it to an implementation
using the host-side communication approach discussed in HPCC FPGA
\autocite{meyer_multi-fpga_2023}. Therefore, we split our work into two
parts.

In the first part in Section~\ref{sec-benchmarks}, we will use synthetic
benchmarks to measure the differences in resource utilization,
communication latency, and throughput for different ACCL configurations
and compare them to the communication approach taken by HPCC FPGA. Also,
we provide models for communication latency and discuss the limitations
and opportunities for different ACCL- and host-based communication
approaches. Moreover, we discuss optimization options for ACCL and its
network stacks for the inter-FPGA network infrastructure of the Noctua~2
supercomputer, which contains one of the largest installations of FPGAs
in the academic HPC domain.

In the second part, in Section~\ref{sec-application}, we port a
multi-FPGA shallow water simulation that operates on unstructured meshes
\autocite{faj2023utbest} and has strong requirements for low-latency
communication to Xilinx FPGAs. We use our findings from
Section~\ref{sec-benchmarks} to identify the best ACCL configuration for
this kind of application and evaluate the scaling behavior of the
application over up to 48 FPGAs compared to a baseline version using
host MPI for communication. In addition, we extend the existing
performance models of the shallow water simulation to also reflect
communication latency to show the effect of high communication latency
on the application scalability.

\section{Related Work}\label{sec-related-work}

This work analyzes ACCL as an inter-FPGA communication framework for
scaling latency-sensitive applications. Several multi-FPGA applications
exist, using either other existing communication frameworks or custom
solutions.

\textcite{fujita_raytracing_2020} accelerate astrophysical simulations
using FPGAs and the inter-FPGA communication framework CIRCUS. They
achieve a high parallel efficiency with a weak scaling scenario on up to
four FPGAs. \textcite{kobayashi_astro_2023} extends the work utilizing
GPUs and FPGAs and using MPI and CIRCUS for communication. The
application showed linear speedups for weak scaling on up to two compute
nodes or up to two FPGAs, respectively. However, the scaling evaluation
for this application is very limited.

\textcite{huthmann_nbody_2019} implemented an N-Body simulation using a
custom circuit-switched network. The implementation showed linear
speedups in strong and weak scaling scenarios over up to 8 FPGAs.
Another N-Body simulation by \textcite{menzel_nbody_2021} uses
SerialLite via the OpenCL Intel External Channel extension to
communicate within a custom circuit-switched network and archives linear
scaling in a strong scaling scenario with up to 24 FPGAs. The
implementation achieves sub-millisecond time step durations.

A shallow water simulation on unstructured meshes by
\textcite{faj2023utbest} similarly makes use of SerialLite within a
custom circuit-switched network. The mesh is therefore partitioned and
distributed over the FPGAs. FPGAs holding neighboring partitions are
directly connected via the custom circuit-switched network for halo
exchanges. Because of that, the number of neighboring partitions is
limited to the number of available ports on the FPGA board. The
implementation is scaled over up to 10 FPGAs and achieves high parallel
efficiency in strong scaling scenarios with a time step duration of only
several microseconds. These small time steps set high requirements on
the communication latency, thus we picked this work for our evaluation.

The HPCC FPGA benchmark suite \autocite{meyer_multi-fpga_2023} contains
multi-FPGA implementations for LU factorization and matrix
transposition. Next to the baseline versions using the naive
communication approach via PCIe and MPI, it also contains optimized
versions of these benchmarks directly utilizing the SerialLite network
stack for communication in custom circuit-switched networks. The LU
factorization does not have tight constraints on communication latency
or throughput since communication latencies can be hidden by computation
for large matrices. The matrix transposition is mainly communication
throughput limited. Both applications were executed on up to 25 FPGAs
showing close to linear speedups for communication in custom
circuit-switched networks.

\section{Synthetic Benchmarking of Communication
Approaches}\label{sec-benchmarks}

\subsection{ACCL Communication Approaches}\label{sec-accl-comm-types}

ACCL offers two communication approaches: \emph{streaming} and
\emph{buffered} communication. \emph{Buffered} communication is similar
to the well-known blocking MPI communication, whereas \emph{streaming}
communication supports the processing of incoming data before the
transmission is complete. This allows further overlapping of
communication and computation. As additional configuration option, ACCL
supports the scheduling of communication from the host side (offering
more flexibility), or directly from FPGA.

\begin{figure*}[tbh]

\begin{minipage}{0.38\linewidth}

\centering{

\includegraphics{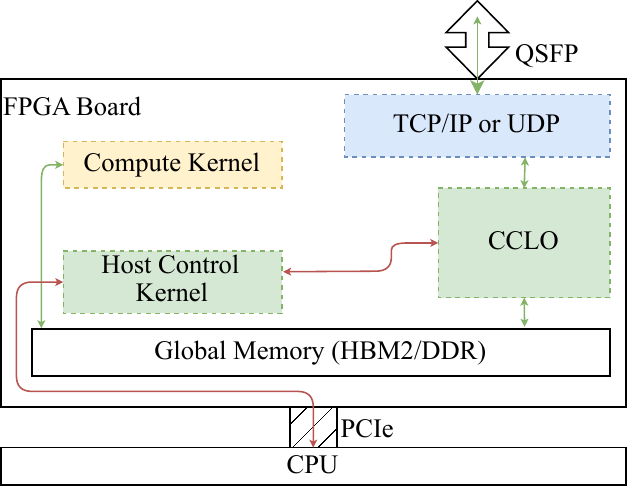}

}

\subcaption{\label{fig-accl-basic}Buffered communication controlled by
the host. Data is exchanged between ACCL and user kernels via global
memory.}

\end{minipage}%
\begin{minipage}{0.03\linewidth}
~\end{minipage}%
\begin{minipage}{0.38\linewidth}

\centering{

\includegraphics{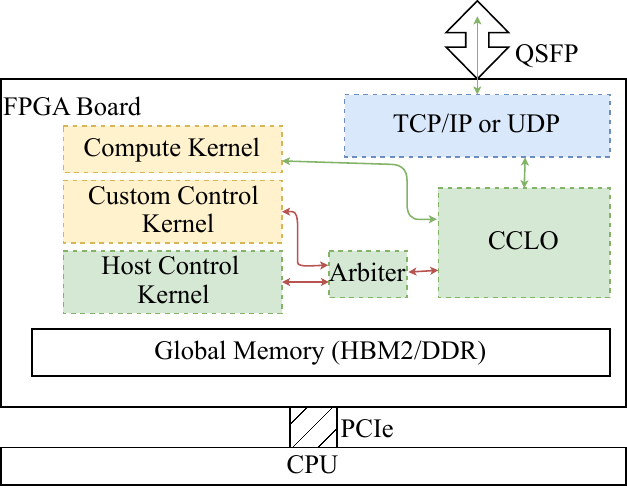}

}

\subcaption{\label{fig-accl-plschedule}Streaming communication
controlled from PL. Data is exchanged between ACCL and user kernels via
AXI streams.}

\end{minipage}%
\begin{minipage}{0.03\linewidth}
~\end{minipage}%
\begin{minipage}{0.20\linewidth}
\includegraphics{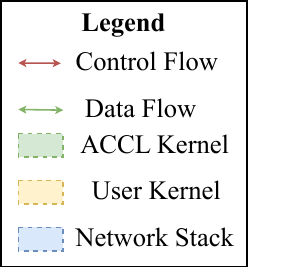}\end{minipage}%

\caption{\label{fig-approaches}Two examples of communication approaches
using the ACCL framework.}

\end{figure*}%

Buffered communication with communication scheduling from the host side
is visualized in Figure~\ref{fig-accl-basic}. Here, ACCL will transfer
data from a buffer in global memory to another buffer in the global
memory of a remote FPGA. The \emph{compute kernel} -- the actual
application implemented on FPGA -- can read the data from this global
memory buffer afterwards. The communication is controlled on the host
via a C++ library. Instead of exchanging data between the \emph{compute
kernel} and the ACCL infrastructure indirectly via global memory, it can
also be directly forwarded using AXI streams. This approach is indicated
in Figure~\ref{fig-accl-plschedule} as a green AXI stream between
\emph{compute kernel} and \emph{CCLO}. A drawback of this approach is,
that the order of incoming messages can not be controlled by the
receiving side because received data is directly forwarded from ACCL to
the AXI stream. If two FPGAs stream a message to the same recipient, the
contents of the message will be forwarded in the order of arrival, which
may also lead to a scattering of messages. The compute kernels need to
be extended to handle these situations.

The other configuration option builds upon the AXI stream interface that
is used to issue commands. In addition to the default \emph{host control
kernel} (Figure~\ref{fig-accl-basic}) that requires a dedicated kernel
invocation for every communication request, it is also possible to
implement \emph{custom control kernels}
(Figure~\ref{fig-accl-plschedule}). ACCL already comes with an API that
can be used from HLS kernels to implement this functionality. A
\emph{custom control kernel} implements the communication pattern
required by a specific \emph{compute kernel} and thus can significantly
reduce the number of required kernel invocations. \emph{Compute kernel}
and \emph{custom control kernel} can also be combined into a single
kernel.

\subsection{Evaluation Infrastructure}\label{sec-infrastructure}

\begin{figure}[tbh]

\centering{

\includegraphics[width=0.7\textwidth,height=\textheight]{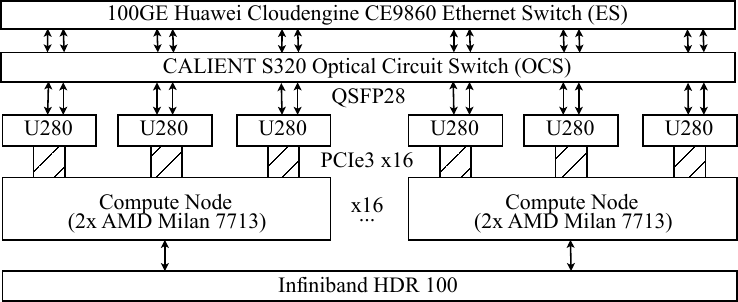}

}

\caption{\label{fig-noctua2}The network infrastructure of the FPGA nodes
within the Noctua 2 cluster. The FPGAs are connected to a dedicated
Ethernet switch via their QSFP28 ports. The compute nodes communicate
via a separate Infiniband network.}

\end{figure}

For our evaluation, we use the FPGA partition of the Noctua 2 cluster at
the Paderborn Center for Parallel Computing (PC\textsuperscript{2}) to
synthesize and execute the benchmarks and applications discussed in this
paper. Noctua 2 contains one of the largest FPGA installations in the
academic HPC domain with a total of 48 Alveo U280 FPGAs distributed over
16 nodes. All FPGAs are connected to an optical switch over their QSFP28
ports as given in Figure~\ref{fig-noctua2}.

The optical switch is protocol agnostic and can be used to physically
connect arbitrary ports of the switch to form direct point-to-point
connections with minimal latency overhead. In addition, a 128-port
ethernet switch is also connected to the optical switch. By configuring
the optical switch, this setup also allows the connection of FPGA ports
to the ethernet switch to form packet-switched networks. We use this
setup in our evaluation to look more deeply into the communication
latencies introduced by packet-switched communication. We used Vitis
2022.2, XRT 2.14, and the shell
\texttt{xilinx\_u280\_gen3x16\_xdma\_1\_202211\_1} for the synthesis and
execution of all applications.

\subsection{Resource Utilization of the Network
Stack}\label{sec-benchmark-resource-utilization}

We used the benchmark \emph{b\_eff} from the HPCC FPGA
\autocite{meyer_multi-fpga_2023} benchmarks suite to evaluate the
resource utilization of ACCL and the network stack on the FPGA.
\emph{b\_eff} is a synthetic benchmark where the FPGAs are arranged in a
(virtual) ring topology to exchange messages. The messages are sent for
a given range of message sizes in a ping-ping fashion between the
neighbors in the ring and can be used to measure the latency and
throughput of the network.

\begin{figure}[tbh]

\centering{

\includegraphics{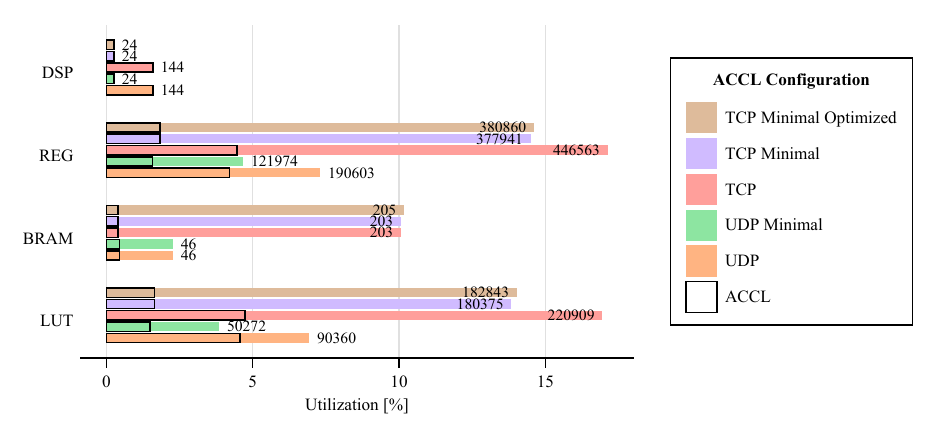}

}

\caption{\label{fig-network-stack-resources}Resource utilization of the
network stack and ACCL on the Alveo U280. The ACCL \emph{Minimal}
versions do not contain the compression and arithmetic plugins. The
resource utilization of ACCL (green boxes in
Figure~\ref{fig-approaches}) is highlighted by black boxes. All further
resources are consumed by the network stack (blue boxes in
Figure~\ref{fig-approaches}).}

\end{figure}%

We implemented \emph{b\_eff} designs with ACCL based on the
implementation presented in Figure~\ref{fig-accl-plschedule}, now
further differentiating between the TCP/IP and the UDP network stack.
The resource utilization for different ACCL configurations is given in
Figure~\ref{fig-network-stack-resources}. Further variants are shown,
because ACCL is extendable by plugins which are included by default and
provide extra functionality for data compression and arithmetic
operations used in collectives, such as reductions. These plugins can be
removed from the ACCL configuration by setting build flags during
synthesis. Since functionality is not needed by \emph{b\_eff}, we also
synthesized a \emph{minimal} ACCL without unused plugins to save
additional resources. Moreover, to optimize the TCP throughput with the
Ethernet switch, we synthesized an \emph{optimized} TCP stack
configuration. We discuss these optimization steps in more detail in
Section~\ref{sec-benchmarks-throughput}. As expected, the TCP network
stack consumes considerably more resources compared to the UDP stack.
Our minimal ACCL version saves more than half of the logic resources and
more than 83\% of DSPs independent of the used network stack.

\subsection{Modelling and Measurement of Throughput and
Latency}\label{sec-benchmarks-throughput}

We executed the \emph{b\_eff} benchmark for the different communication
approaches discussed in Section~\ref{sec-accl-comm-types} on two FPGAs.
For the first experiment, we configured the optical switch to create a
direct connection between the FPGAs -- bypassing the ethernet switch.
Furthermore, the HPCC FPGA version from the original benchmark without
ACCL was used to retrieve data for a purely CPU-based baseline. The two
FPGAs are located on different nodes, such that data transfers of the
baseline version use the Infiniband network of the hosts via MPI.

\begin{figure}[tbh]

\centering{

\includegraphics{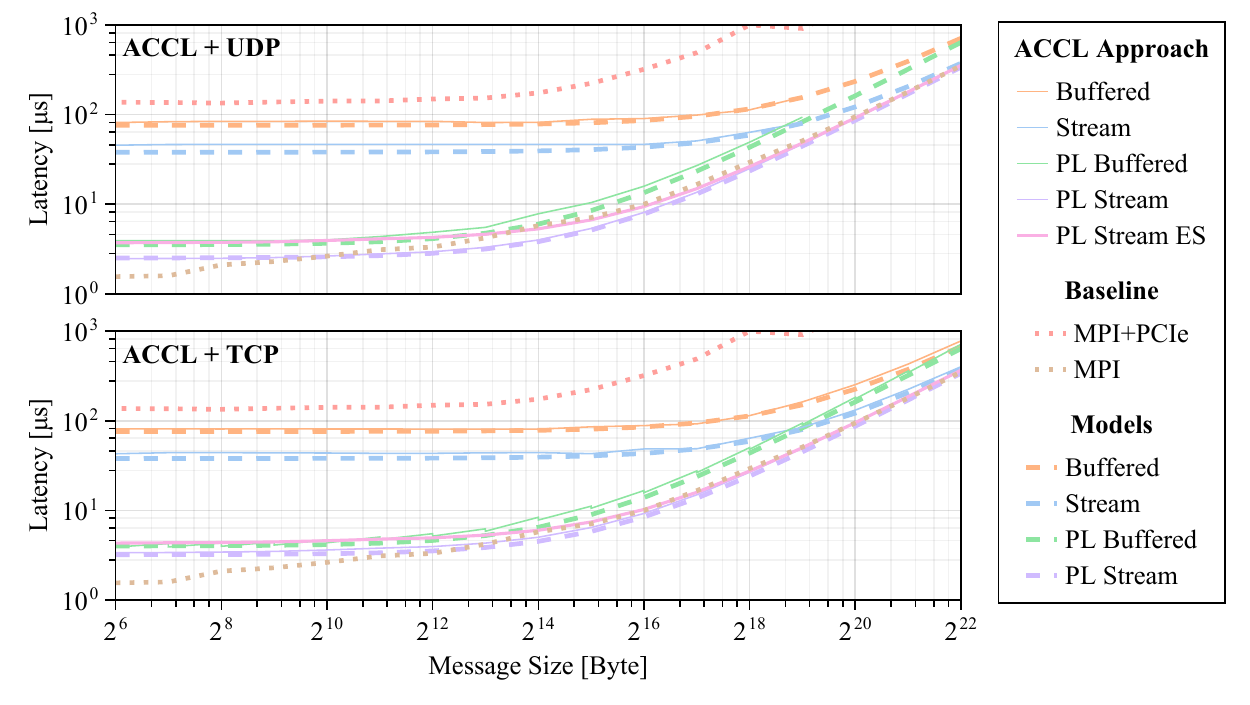}

}

\caption{\label{fig-beff-latencies}Full-duplex communication latencies
using ACCL UDP and TCP network stacks and the discussed communication
approaches. The latencies for host-side scheduling are modeled for
buffered and streaming communication. Communication via the ethernet
switch is marked with ES in the legend. All other measurements are done
with directly connected FPGAs.}

\end{figure}%

The communication latencies over the message size are given for
\emph{streaming} and \emph{buffered} communication in
Figure~\ref{fig-beff-latencies}, each in combination with communication
scheduling from the host and with a custom control kernel from the FPGA
(denoted here as \emph{PL}). We also integrated a performance model for
the different approaches as dashed lines in the plot. As expected, the
baseline communication approach shows the highest communication latency
over all message sizes with latencies of more than 120μs for 64 Byte
messages. Using the buffered ACCL communication scheduled from the host
side, the major limitation for the latency becomes the kernel scheduling
time. The ACCL host control kernel needs to be executed two times for
sending and receiving a message. We measured around 30µs of latency for
a single kernel invocation through the used XRT runtime. In contrast,
messages scheduled directly from \emph{PL} reach latencies below 3µs.

We modeled the latency for buffered and streamed communication using the
theoretical peak throughput of the involved links as well as our
measured kernel start overheads. For buffered communication, this
results in the model given in Equation~\ref{eq-host-buffered-model},
where \(l_k\) is the time required to schedule a command to ACCL,
\(l_{m}\) the latency to copy the message from the receive buffer in
global memory to the destination buffer on the receiving FPGA, and
\(l_{c}\) is the latency of the communication link. For host scheduled
communication, \(l_{k}\) equals the kernel invocation latency, whereas
for PL scheduling it is reduced to a fraction of microseconds since it
only represents the time required by ACCL to process the command.

\begin{equation}\phantomsection\label{eq-host-buffered-model}{
2 \cdot l_k + l_{m} + l_{c}
}\end{equation}

For streaming, the model simplifies to \(l_k + l_c\), because only a
single kernel invocation is required per transmission. Also, there is no
copy operation required since the data is directly passed to the AXI
stream of the user kernel. While \(l_{k}\) is a constant overhead per
ACCL command, \(l_{c}\) and \(l_{m}\) depend on the message size and
overtake the equation for large message sizes. The additional copy
operation required in buffered communication thus leads to a reduced
theoretical peak throughput of
\((14\text{GB/s}^{-1}+12.5\text{GB/s}^{-1})^{-1} = 6.6 \text{GB/s}\) --
only slightly more than half of the peak throughput of the communication
link.

Additionally, we compared the communication latency of the network
configuration with direct optical links to the latency of connections
via the ethernet switch. Overall, the ethernet switch adds around 1μs of
latency for both network stacks resulting in latencies of \(2.5\) to 5
µs for small 64 Byte messages. For the TCP stack, the throughput was at
first considerably reduced when using the Ethernet switch. Because of
the increased communication latency, the sending side has to stop
transmission and wait for acknowledgments. In our optimized TCP
implementation, we enabled window scaling to overcome this issue, with a
minor impact on resource consumption as shown in
Section~\ref{sec-benchmark-resource-utilization}. Moreover, we reduced
the overall protocol overhead by enabling jumbo frames on the Ethernet
switch and increasing the maximum segment size for the TCP stack and the
maximum packet size in the ACCL firmware accordingly. These changes
increased the throughput for large messages from initially 8.5~GB/s with
the TCP stack to 12.3~GB/s for both network stacks while having no
measurable impact on the latency for small messages. The measurements
via the ethernet switch for the optimized TCP and UDP stack are also
given in Figure~\ref{fig-beff-latencies} and annotated with \emph{ES}.

\section{Acceleration of Shallow Water Simulation using
ACCL}\label{sec-application}

\subsection{Implementation}\label{implementation}

The evaluation of ACCL using synthetic benchmarks showed that low
latency communication in the order of a few µs is possible. We applied
the lessons learned to a full FPGA accelerated HPC application: A
discontinuous Galerkin shallow water simulation on unstructured meshes
which was originally implemented for Intel FPGAs
\autocite{kenter2021utbest} and was further extended for multi-FPGA
execution in custom circuit-switched communication networks
\autocite{faj2023utbest}.

\begin{figure}

\begin{minipage}{0.49\linewidth}

\begin{figure}[H]

\centering{

\includegraphics[width=0.8\textwidth,height=\textheight]{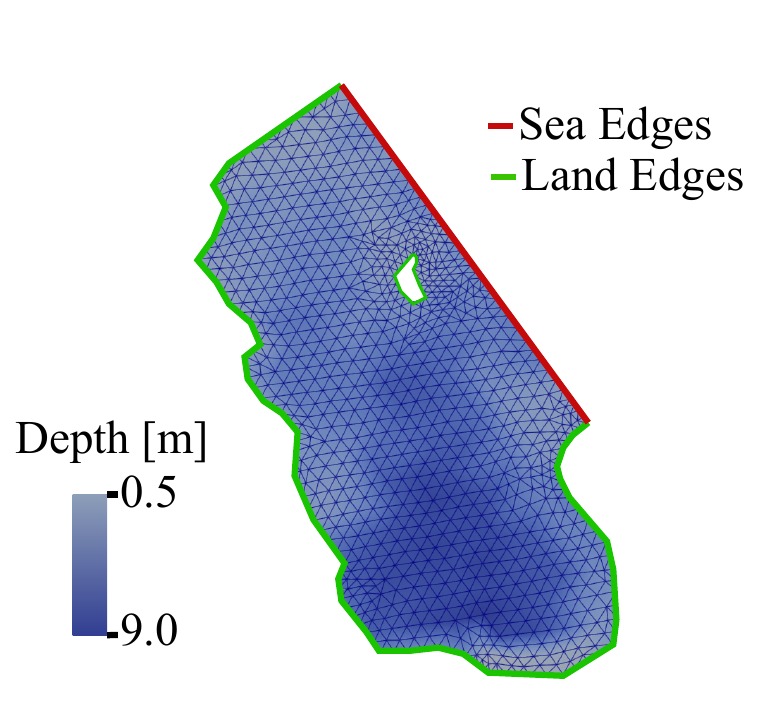}

}

\caption{\label{fig-utbest-visual}Computational mesh consisting of 1696
elements \autocite{faj2023utbest}. The boundary edges represent the
coastline (land edges) and the border to the open sea (sea edges).}

\end{figure}%

\end{minipage}%
\begin{minipage}{0.02\linewidth}
~\end{minipage}%
\begin{minipage}{0.49\linewidth}

\begin{figure}[H]

\centering{

\includegraphics[width=0.6\textwidth,height=\textheight]{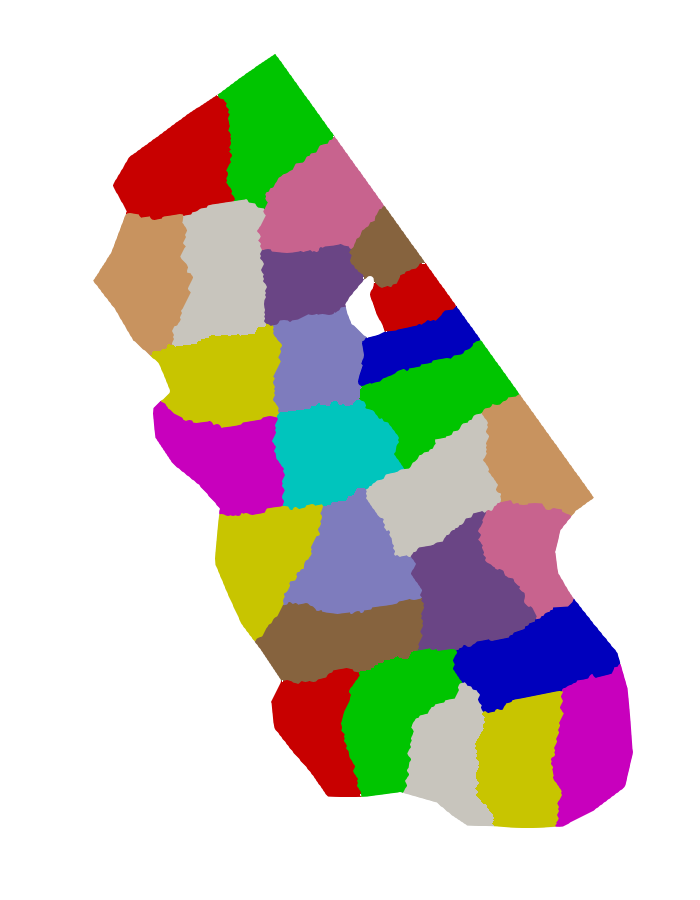}

}

\caption{\label{fig-utbest-partitioning}The mesh is partitioned as
indicated by the colors. Data has to be exchanged between neighboring
partitions in every simulation time step.}

\end{figure}%

\end{minipage}%

\end{figure}%

The tidal flow of the bight of Abaco on Bahamas islands is used as a
simulation scenario as given in Figure~\ref{fig-utbest-visual}. The
water surface of the bay is represented by an unstructured mesh. The
borders of the mesh are either land or sea edges which have to be
handled differently in the simulation. For the execution over multiple
FPGAs, the mesh is partitioned into sub-meshes as visualized in
Figure~\ref{fig-utbest-partitioning}. The mesh partitions are
distributed among the FPGAs, such that every FPGA handles exactly one
partition. In each simulation step, the halo around the partition edges
has to be exchanged between neighboring FPGAs using point-to-point
communication. The designs in \autocite{kenter2021utbest} and
\autocite{faj2023utbest} support three types of polynomial discontinuous
Galerkin discretizations, however, it has been shown by Faj et al.
\autocite{faj2023utbest} that the requirements for communication latency
are very similar for all three types of discretization. Thus, we will
only focus on the piecewise constant discretization in our evaluation.

\begin{figure}[tbh]

\centering{

\includegraphics[width=0.9\textwidth,height=\textheight]{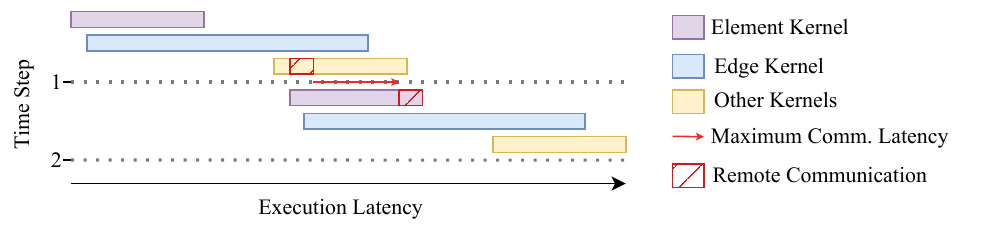}

}

\caption{\label{fig-utbest-latency}Dataflow schematic for shallow water
simulation. Full overlap of compute kernels across simulation time
steps. The maximum communication latency is indicated by the red arrow.
If communication takes longer, the compute pipeline will stall until
data is received. Simplified dataflow based on \textcite{faj2023utbest},
Fig. 5}

\end{figure}%

Figure~\ref{fig-utbest-latency} shows the dataflow graph over two
simulation time steps for the compute pipeline in a single FPGA. All
simulation data is loaded into the local memory of the FPGA at the
beginning of the simulation so global memory accesses are only used
during simulation to write back intermediate results. The \emph{element
kernel} updates all entities of the unstructured mesh element-wise. The
L\^{}2 projection is directly forwarded to the \emph{edge kernel} which
will update all boundary and outer edges of the unstructured mesh.
Afterwards, the boundary edges between partitions of the unstructured
mesh will be sent within the \emph{other kernels} to the remote FPGAs
via ACCL. This data will be received at the end of the execution of the
\emph{element kernel} in the next time step. To prevent pipeline stalls,
the data has to arrive at the remote FPGA within the latency indicated
by the red arrow which is typically a few thousand clock cycles. The
overall latency between sending and receiving boundary elements depends
mainly on the number of core elements that will be updated by the
\emph{element kernel} in between. Core elements are elements that are
not located on a border of the local partition and which do not require
any data from remote FPGAs.

\begin{figure}[tbh]

\centering{

\includegraphics{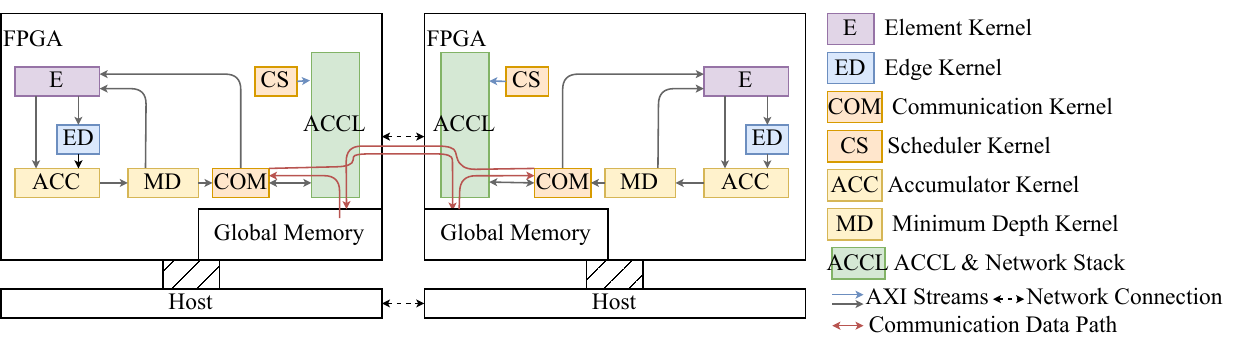}

}

\caption{\label{fig-utbest-accl-compact}Remotely partitioned FPGA
designs with two processing pipelines distributed over two FPGAs with
ACCL communication via AXI streams and communication scheduling in PL.}

\end{figure}

The data is streamed through the kernels element-wise as given in
Figure~\ref{fig-utbest-accl-compact}. The mesh partition will be updated
on the local FPGA and all boundary elements are forwarded to a
\emph{communication kernel} which has the task of passing this data to
the compute pipeline on the remote FPGA. In the baseline implementation,
the \emph{communication kernel} is invoked by the host for every
simulation step. It will write all received data into a buffer in global
memory and finish execution after all elements for one simulation step
are written to notify the host that the data is ready for sending via
MPI. The remote host calls the \emph{communication kernel} again to read
the data from global memory to the \emph{element kernel} for the next
simulation step.

For the ACCL-enabled version, the \emph{communication kernel} converts
the simulation data into a generic 512-bit AXI stream used to directly
pass the data into the ACCL communication stack. This way, the actual
simulation loop stays unchanged. In addition, we use a communication
scheduler kernel to issue the send and receive commands directly from
PL. This massively reduces the number of kernel invocations from the
host side.

For a high number of partitions, there will be more than one neighboring
partition and the \emph{element kernel} expects the remote elements from
the communication kernel in a predefined order. As a result, the
incoming data has to be reordered on the receiver side before it can be
passed onto the simulation pipeline. Instead of creating our own logic
for this task, we make use of ACCL's buffered communication feature.
Therefore, we buffer received data in global memory first and move the
data from the global memory into the AXI stream using the \texttt{recv}
primitive as indicated by the red arrows in the figure. In the end, the
resulting communication scheme is a mixture of streaming communication
on the sending side and buffered communication on the receiving side.
This approach slightly increases the communication latency because data
has to go through the global memory instead of passing it directly to
the communication kernel. As a main advantage, we have no strict upper
limit for the number of neighboring partitions since we can configure
the number of receive buffers during runtime.

\subsection{Performance Model}\label{sec-performance-model}

The port of the simulation to Vitis-compatible C++ code was possible
without major changes in the dataflow characteristics. This also means,
that the performance models for the simulation proposed by
\textcite{faj2023utbest} also hold for this implementation. However,
their throughput model does not consider the communication latency,
which is important for strong scaling scenarios and small local
partition sizes, since the calculations on the core elements may not be
sufficient to hide the communication latency. Because of this, we
extended the existing throughput model as given in
Equation~\ref{eq-throughput} with the communication latency \(L_{comm}\)
representing the latency for the FPGA with the highest number of
neighbors according to the partition scheme and the largest number of
sent or received elements per simulation time step.

\begin{equation}\phantomsection\label{eq-throughput}{
throughput = f \cdot \frac{FLOP_{total}}{max(E_{core} + D_{ext}, L_{comm}) + E_{send} + E_{recv} + L_{pipe}}
}\end{equation}

Additionally, we model the communication latency based on our latency
measurements in Section~\ref{sec-benchmarks}, our ACCL latency models,
as well as mesh partitioning information as given in
Equation~\ref{eq-comm-latency}. As described in the previous section, to
receive the data, it has to be read from a buffer in global memory with
a latency of \(l_{m}\). The maximum number of neighbors \(N_{max}\) for
a partition scheme has a major effect on this read latency because the
read commands have to be scheduled for every neighbor. This adds
\(l_{m}\) to the overall communication latency for every neighbor as
given in Equation~\ref{eq-host-buffered-model}. The latency to process
the commands in ACCL \(l_{k}\) has to be added for every send and
receive command.

\begin{equation}\phantomsection\label{eq-comm-latency}{
L_{comm} =  \frac{E_{send} + E_{recv} + 2 \cdot N_{max} \cdot l_{k} + N_{max} \cdot l_{m}}{f} + L_{pingping}
}\end{equation}

Also, the overall latency of the communication link has to be
considered, introducing another latency \(L_{pingping}\), which is the
ping-ping latency of the largest message exchanged among neighbors. For
the total number of floating-point operations \(FLOPS_{total}\) we use
the simplified model \(FLOPS_{total} = FLOP_{sum} \cdot E_{total}\)
without the additional operations required to calculate the projection
on the received elements. Instead, we calculate the FLOPs based on the
total number of elements in the mesh \(E_{total}\) and the number of
floating point operations per element \(FLOPS_{sum}\), independent of
the partitioning to make scaling experiments better comparable.

\subsection{Evaluation}\label{evaluation}

We synthesized the base and ACCL version of our shallow water simulation
for the same infrastructure described in
Section~\ref{sec-infrastructure}. For our application, we use the UDP
minimal and TCP minimal optimized configurations of ACCL as described in
Section~\ref{sec-benchmark-resource-utilization}. The resulting resource
utilization is given in Table~\ref{tbl-utbest-resource-usage} for the
used configurations. The increased resource utilization compared to the
base version closely reflects the resource usage of the ACCL stack. The
implementation supports setting the maximum number of elements per
partition, which mainly affects the BRAM and URAM utilization since the
whole partition is stored in local memory. The baseline and UDP versions
are synthesized with a partition size of 8192 elements where larger
sizes led to routing congestion because the URAMs holding the local
partition can not be easily distributed across multiple
sliced logic regions (SLRs). For the ACCL TCP stack, we were only able
to synthesize designs with half the partition size, i.e.~4096 elements.
Larger local partitions also failed because of routing congestion.

\begingroup
\scriptsize

\begin{longtable}[]{@{}
  >{\raggedright\arraybackslash}p{(\columnwidth - 14\tabcolsep) * \real{0.1600}}
  >{\raggedleft\arraybackslash}p{(\columnwidth - 14\tabcolsep) * \real{0.1300}}
  >{\raggedleft\arraybackslash}p{(\columnwidth - 14\tabcolsep) * \real{0.1300}}
  >{\raggedleft\arraybackslash}p{(\columnwidth - 14\tabcolsep) * \real{0.1200}}
  >{\raggedleft\arraybackslash}p{(\columnwidth - 14\tabcolsep) * \real{0.1200}}
  >{\raggedleft\arraybackslash}p{(\columnwidth - 14\tabcolsep) * \real{0.1200}}
  >{\raggedleft\arraybackslash}p{(\columnwidth - 14\tabcolsep) * \real{0.1000}}
  >{\raggedleft\arraybackslash}p{(\columnwidth - 14\tabcolsep) * \real{0.1200}}@{}}

\caption{\label{tbl-utbest-resource-usage}Resource utilization of the
shallow water simulation}

\tabularnewline

\toprule\noalign{}
\begin{minipage}[b]{\linewidth}\raggedright
\textbf{Configuration}
\end{minipage} & \begin{minipage}[b]{\linewidth}\raggedleft
\textbf{LUTs}
\end{minipage} & \begin{minipage}[b]{\linewidth}\raggedleft
\textbf{Registers}
\end{minipage} & \begin{minipage}[b]{\linewidth}\raggedleft
\textbf{BRAM}
\end{minipage} & \begin{minipage}[b]{\linewidth}\raggedleft
\textbf{URAM}
\end{minipage} & \begin{minipage}[b]{\linewidth}\raggedleft
\textbf{DSPs}
\end{minipage} & \begin{minipage}[b]{\linewidth}\raggedleft
\textbf{Freq. {[}MHz{]}}
\end{minipage} & \begin{minipage}[b]{\linewidth}\raggedleft
\textbf{Synth. Time {[}h{]}}
\end{minipage} \\
\midrule\noalign{}
\endhead
\bottomrule\noalign{}
\endlastfoot
{\textbf{Base}} & \begin{minipage}[t]{\linewidth}\raggedleft
{126,646}\\
{(9.7\%)}\strut
\end{minipage} & \begin{minipage}[t]{\linewidth}\raggedleft
{182,015}\\
{(7.0\%)}\strut
\end{minipage} & \begin{minipage}[t]{\linewidth}\raggedleft
{265}\\
{(13.1\%)}\strut
\end{minipage} & \begin{minipage}[t]{\linewidth}\raggedleft
{188}\\
{(19.6\%)}\strut
\end{minipage} & \begin{minipage}[t]{\linewidth}\raggedleft
{1,218}\\
{(13.5\%)}\strut
\end{minipage} & {256} & {4.1} \\
{\textbf{ACCL UDP}} & \begin{minipage}[t]{\linewidth}\raggedleft
{176,884}\\
{(13.6\%)}\strut
\end{minipage} & \begin{minipage}[t]{\linewidth}\raggedleft
{305,381}\\
{(11.7\%)}\strut
\end{minipage} & \begin{minipage}[t]{\linewidth}\raggedleft
{312}\\
{(15.5\%)}\strut
\end{minipage} & \begin{minipage}[t]{\linewidth}\raggedleft
{188}\\
{(19.6\%)}\strut
\end{minipage} & \begin{minipage}[t]{\linewidth}\raggedleft
{1,242}\\
{(13.8\%)}\strut
\end{minipage} & {274} & {5.2} \\
{\textbf{ACCL TCP}} & \begin{minipage}[t]{\linewidth}\raggedleft
{334,225}\\
{(25.6\%)}\strut
\end{minipage} & \begin{minipage}[t]{\linewidth}\raggedleft
{586,847}\\
{(22.5\%)}\strut
\end{minipage} & \begin{minipage}[t]{\linewidth}\raggedleft
{344}\\
{(17.1\%)}\strut
\end{minipage} & \begin{minipage}[t]{\linewidth}\raggedleft
{101}\\
{(10.5\%)}\strut
\end{minipage} & \begin{minipage}[t]{\linewidth}\raggedleft
{1,242}\\
{(13.8\%)}\strut
\end{minipage} & {252} & {15.1} \\

\end{longtable}

\endgroup

\begin{figure}[tbh]

\centering{

\includegraphics{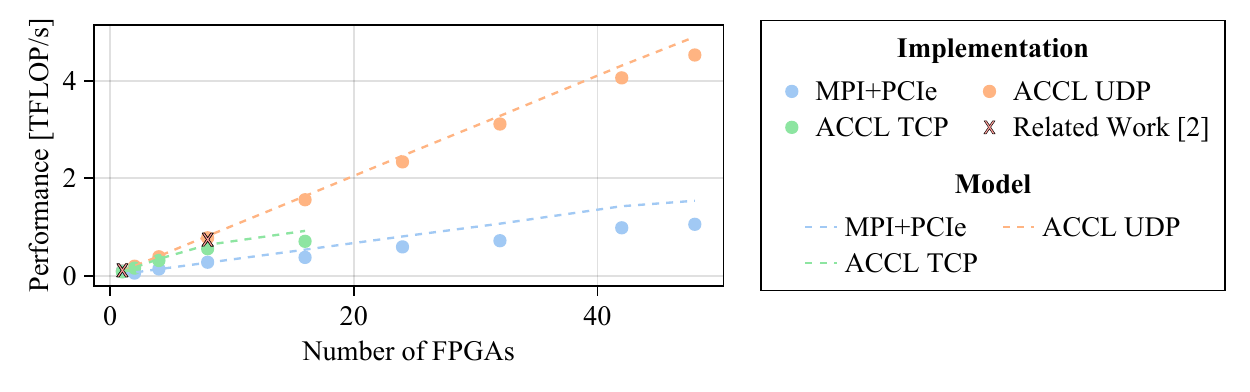}

}

\caption{\label{fig-utbest-execution-times}Execution times for the weak
scaling scenario with \textasciitilde6000 elements per FPGA. Due to a
known issue in the current version of the TCP stack, we did not evaluate
this configuration beyond 16 devices.}

\end{figure}%

\begin{figure}[tbh]

\centering{

\includegraphics{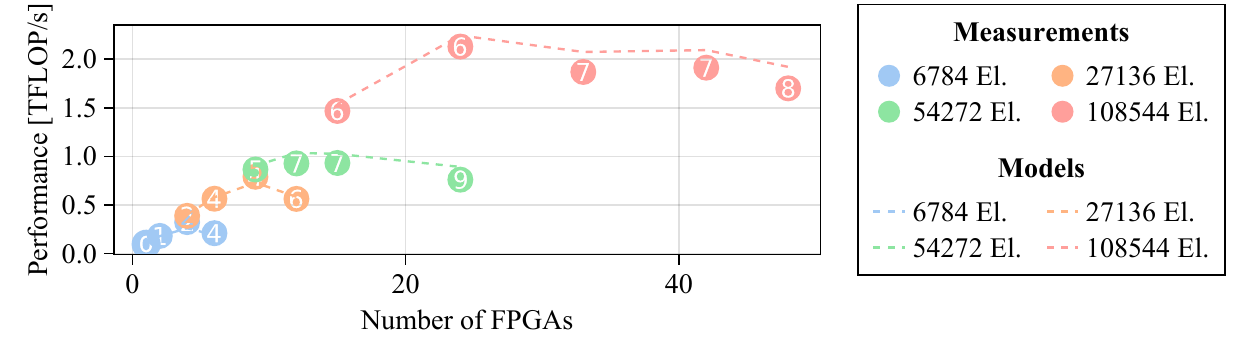}

}

\caption{\label{fig-utbest-strong-scaling}Strong scaling scenario with
selected mesh sizes with ACCL and UDP stack. The numbers in the plot
represent the maximum number of neighbors.}

\end{figure}%

We first executed a weak scaling experiment with the three design
variants of the shallow water simulation. The resulting performance
compared to the modified performance model is given in
Figure~\ref{fig-utbest-execution-times}. We used increasing mesh sizes
with up to 312,000 elements to keep the number of elements per partition
between 6,000-7,000 elements. The base version annotated as
\emph{MPI+PCIe} first shows a reduced performance when scaling from one
partition to two partitions. When executed only on one partition, no
communication is required, eliminating compute pipeline stalls.
Measurements with our synthetic benchmark given in
Figure~\ref{fig-beff-latencies} showed an expected latency of 100-120µs
for small messages of multiple KB size. This is the expected size of the
halo exchange messages sent by the shallow water simulation. The
simulation processes one element per clock cycle and sufficient core
elements are required to hide the communication latency as expressed in
the \emph{max} term in Equation~\ref{eq-throughput}. Based on the kernel
frequency of the synthesized bitstream, the pipeline could process
around 25,000 to 30,000 elements in this time frame, so for the given
partitions, the pipeline stalls approximately 75-80\% of the execution
time. With their improved communication latency, the ACCL designs with
UDP and TCP stack can respectively avoid or reduce such stalls, which
leads to higher performance and better scalability, to up to 4.5
measured TFLOPs on 48 FPGAs with the ACCL UDP design.

Furthermore, we executed strong scaling experiments with selected mesh
sizes given in Figure~\ref{fig-utbest-strong-scaling}, which
additionally depict the maximum number of communication neighbors. The
results show that this number has a high impact on the overall
performance, because of the additional latency introduced by command
scheduling and global memory. When the local partitions become too small
to cover the communication latency, there is no performance improvement
by adding further FPGAs. Indeed, the overall performance can even
degrade, because further partitioning of the mesh may introduce a higher
maximum number of neighbors, which in turn further increases the
communication latency. This can be clearly observed for the 108K element
mesh, where additional neighbors result in a step-wise decrease in
performance. The original implementation of \autocite{faj2023utbest} is
limited to a maximum of four neighbors because of the number of QSFP
ports installed on the FPGA board and thus was limited to at most 10
FPGAs for the topology used in their and our experiments. Our ACCL
implementation overcomes this limitation, but we see that in strong
scaling scenarios, larger local partition sizes or custom message
reordering in local memory would be required to hide the communication
overheads introduced by the increasing number of neighbors.

\section{Conclusion}\label{conclusion}

In this work, we modeled the communication latency with ACCL for
buffered and streamed communication and showed, that buffered
communication leads to latency and throughput degradation because of
additional copy operations. Also, communication scheduling from PL was
shown to drastically improve communication latency because of a reduced
number of kernel invocations. Based on our ACCL evaluation, we ported a
multi-FPGA shallow water simulation to Xilinx FPGAs and extended it with
communication via ACCL. The scaling experiments showed linear speedups
in weak scaling scenarios with all 48 FPGAs of the Noctua 2 partition
and the same performance per partition, but with much better scalability
than an implementation using custom circuit-switched networks. This
makes ACCL and packet-switched network infrastructures good candidates
for scaling communication-latency-sensitive multi-FPGA applications.

\subsubsection*{Acknowledgements}\label{acknowledgements}
\addcontentsline{toc}{subsubsection}{Acknowledgements}

The authors gratefully acknowledge the computing time provided to them
on the high-performance computer Noctua 2 at the NHR Center PC2. These
are funded by the Federal Ministry of Education and Research and the
state governments participating based on the resolutions of the GWK for
the national high-performance computing at universities
(\texttt{www.nhr-verein.de/unsere-partner}). Also, they thank Markus
Büttner and Jonathan Schmalfuss from the University of Bayreuth for the
help with mesh generation for the shallow water simulation.

This work was supported in part by AMD under the Heterogeneous
Accelerated Compute Clusters (HACC) program. AMD, the AMD Arrow logo,
Alveo™, Vitis™, Xilinx and combinations thereof are trademarks of
Advanced Micro Devices, Inc.~Other product names used in this
publication are for identification purposes only and may be trademarks
of their respective companies.

\printbibliography[title=References]

\end{document}